\documentclass[runningheads]{llncs}
\usepackage{graphicx}
\usepackage{amsmath}
\usepackage{amssymb}
\usepackage{color}
\usepackage{listings}

\usepackage{listings, xcolor}

\usepackage[scaled=0.8]{DejaVuSansMono}
\usepackage[T1]{fontenc}

\definecolor{verylightgray}{rgb}{.97,.97,.97}

\lstdefinelanguage{Solidity}{
keywords=[1]{anonymous, assembly, assert, balance, break, call, callcode,
case,
catch, class, constant, continue, contract, debugger, default, delegatecall,
delete, do, else, event, export, external, false, finally, for, function, gas,
if, implements, import, in, indexed, instanceof, interface, internal, is,
length, library, log0, log1, log2, log3, log4, memory, modifier, new, payable,
pragma, private, protected, public, pure, push, require, return, returns,
revert, selfdestruct, send, storage, struct, suicide, super, switch, then,
this, throw, transfer, true, try, typeof, using, view, while, with,
addmod, ecrecover, keccak256, mulmod, ripemd160, sha256, sha3}, 
	keywordstyle=[1]\color{darkgray}\bfseries,
keywords=[2]{address, bool, byte, bytes, bytes1, bytes2, bytes3, bytes4,
bytes5, bytes6, bytes7, bytes8, bytes9, bytes10, bytes11, bytes12, bytes13,
bytes14, bytes15, bytes16, bytes17, bytes18, bytes19, bytes20, bytes21,
bytes22, bytes23, bytes24, bytes25, bytes26, bytes27, bytes28, bytes29,
bytes30, bytes31, bytes32, enum, int, int8, int16, int24, int32, int40, int48,
int56, int64, int72, int80, int88, int96, int104, int112, int120, int128,
int136, int144, int152, int160, int168, int176, int184, int192, int200,
int208,
int216, int224, int232, int240, int248, int256, mapping, string, uint, uint8,
uint16, uint24, uint32, uint40, uint48, uint56, uint64, uint72, uint80,
uint88,
uint96, uint104, uint112, uint120, uint128, uint136, uint144, uint152,
uint160,
uint168, uint176, uint184, uint192, uint200, uint208, uint216, uint224,
uint232, uint240, uint248, uint256, var, void, ether, finney, szabo, wei,
days,
hours, minutes, seconds, weeks, years},	
	keywordstyle=[2]\color{teal},
keywords=[3]{block, blockhash, coinbase, difficulty, gaslimit, number,
timestamp, msg, value, data, gas, sender, sig, value, now, tx, gasprice, origin},	
	keywordstyle=[3]\color{violet},
	identifierstyle=\color{black},
	sensitive=false,
	comment=[l]{//},
	morecomment=[s]{/*}{*/},
	morestring=[b]',
	morestring=[b]",
	escapechar=~
}

\lstdefinelanguage{Flint}{
keywords=[1]{associatedtype, protocol, struct, private, typeclass, break,
case, return,
catch, continue, default, defer, do, else, fallthrough, for, guard, if,
in,
repeat, return, switch, throw, try, where, while, implicit, import, assert, send, public,
init, func, let, var, false, true, nil, contract, @payable, @privileged, any, self},
	keywordstyle=[1]\color{darkgray}\bfseries,
keywords=[2]{String, Address, Int, Bool, Void, Wei, Event, Ether, State},
	keywordstyle=[2]\color{teal},
keywords=[3]{mutating},
    keywordstyle=[3]\color{darkgray}\bfseries,
	identifierstyle=\color{black},
	sensitive=false,
	comment=[l]{//},
	morecomment=[s]{/*}{*/},
	commentstyle=\color{gray}\ttfamily,
	stringstyle=\color{red}\ttfamily,
	morestring=[b]',
	morestring=[b]",
	escapechar=~
}

\lstdefinelanguage{bash}{
keywords=[1]{Error},
	keywordstyle=[1]\color{black}\bfseries,
keywords=[2]{Warning},
	keywordstyle=[2]\color{black}\bfseries,
	identifierstyle=\color{black},
	sensitive=false,
	comment=[l]{//},
	morecomment=[s]{/*}{*/},
	commentstyle=\color{gray}\ttfamily,
	stringstyle=\color{black}\ttfamily,
	morestring=[b]',
	morestring=[b]",
	escapechar=~
}

\lstdefinestyle{bash}{
  	backgroundcolor=\color{white},
  	frame=single,
	extendedchars=true,
	showstringspaces=false,
	showspaces=false,
	numbers=none,
	numberstyle=\footnotesize,
	numbersep=9pt,
	tabsize=2,
	breaklines=true,
	showtabs=false,
	captionpos=b,
  	literate={~} {$\sim$}{1}
}

\newcommand{\forget}[1]{ }

\begin{document}

\title{Flint for Safer Smart Contracts}

\author{Franklin Schrans\inst{2} \and
Daniel Hails\inst{1} \and 
Alexander Harkness\inst{1} \and
Sophia Drossopoulou\inst{1} \and
Susan Eisenbach\inst{1} }
\authorrunning{F. Schrans et al.}

\institute{Imperial College London, London SW7 2AZ, UK\\
\email{\{susan, sd\}@ic.ac.uk}
 \and 
Franklin Schrans's contributed while a student at Imperial College.\\
\email{fr@nklinschrans.com} }

\maketitle              

\begin{abstract}
The Ethereum blockchain platform supports the execution of decentralised applications or smart contracts. These typically hold and transfer digital currency to other parties on the platform; however, they have been subject to numerous attacks due to the unintentional introduction of bugs. Over a billion dollars worth of currency has been stolen since its release in July 2015. As smart contracts cannot be updated after deployment, it is imperative that the programming language supports the development of robust contracts.

We propose Flint, a new statically-typed programming language specifically designed for writing robust smart contracts. Flint's features enforce the writing of safe and predictable code. To encourage good practices, we introduce \textit{protection blocks}.  Protection blocks restrict who can run code and when (using typestate) it can be executed. To prevent vulnerabilities relating to the unintentional loss of currency, Flint \textit{Asset} traits provide safe atomic operations, ensuring the state of contracts is always consistent. Writes to state are restricted,  simplifying reasoning about smart contracts. 

\end{abstract}

\section{Introduction}

The Ethereum Virtual Machine\cite{ethereumyellowpaper,ethereumwhitepaper} (EVM) is an open network supporting decentralised execution of programs, known as \textit{smart contracts}.
The EVM is similar to a stateful web service, but instead of being executed by computers controlled by an organisation it is deployed to its nodes (or \textit{miners}).
Smart contracts are held in an append-only data structure a \textit{blockchain}  composed of \textit{blocks}, allowing miners to maintain a consistent view of the network's state. Cryptographic schemes ensure old blocks cannot be modified. Miners select which transaction to process from their transaction pool.

Users can interact with a smart contract by calling the functions it exposes. Function calls are executed by miners, which maintain the state of each smart contract and are paid for processing transactions. Ethereum users and smart contracts can exchange a digital currency known as \textit{Ether} whose smallest denomination is the \textit{Wei} ($10^{-18}$ Ether).  Users also use Ether to purchase \textit{gas}, required to pay for computational costs when executing transactions. 

Smart contracts implement self-managed agreements, enforced autonomously. The source code of a smart contract is available, and {\emph cannot} be changed after deployment. Individuals who interact with smart contracts trust the correct execution of the code rather than reprogrammable machines controlled by a single authority. Smart contracts have been used to implement auctions, votes\cite{soliditydocumentation}, and sub-currencies\cite{openzep} for crowdfunding purposes. Voters do not have to place their trust in the integrity of an electoral organisation when the votes are counted using a smart contract.

Not being able to update a smart contract's code after deployment requires it to be bug free. Attackers have found vulnerabilities in smart contracts allowing the redirection of Ether funds to their personal Ethereum account. Attacks against \textsc{TheDAO}\cite{philip_daian} and the Multi-sig Wallet smart contracts\cite{parity1,parity2} have accumulated losses of over a billion dollars worth of Ether. 

The primary programming language used to write smart contracts, \textit{Solidity}\cite{soliditydocumentation}, is expressive and introduces features designed for smart contract programming. However, Solidity supports a variety of unsafe patterns\cite{atzei2017survey} which makes it difficult for analysis tools\cite{makingsmartcontractssmarter,Mythril} and programmers to find all vulnerabilities. Solidity has few built-in security mechanisms and even worse, vulnerabilities are easily introduced because of simple programming mistakes, such as forgetting a modifier. Others are harder to notice, such as implicit integer overflows, or discarding the return value of sensitive functions.

For traditional problems, languages such as Java\cite{java}, Haskell\cite{jones2003haskell}, Swift\cite{swift}, Rust\cite{rust}, and Pony\cite{pony} leverage years of research in programming languages to prevent the writing of unsafe code. In contrast, multiple programming languages\cite{soliditydocumentation,lll,mutan,serpent,vyper} for writing smart contracts, including Solidity, have attempted to mimic languages such as JavaScript\cite{flanagan2006javascript} and Python\cite{python}, without providing additional safety mechanisms for Ethereum's unique programming model. 

Smart contracts introduce new challenges, which we address in our statically-typed programming language Flint\footnote{Flint was made open source on GitHub\cite{flintgithub} in April 2018 under the MIT license.}, specifically designed for writing smart contracts. By identifying challenges and learning from past vulnerabilities, Flint's features facilitate the development of robust code, and make it more difficult and unnatural to write vulnerable contracts. We highlight the features that should aid in the development of robust code:

\begin{enumerate}
  \item {\bf{Protection Blocks:}} Smart contracts often carry out sensitive operations which need to be protected from unauthorised calls. A call can be unauthorised because the caller shouldn't be allowed to make the call or because  (using \textit{typestate}\cite{fickle}) the contract isn't a valid state to be executed (e.g. until you join a club you cannot participate in its activities).  Flint requires programmers to systematically think about which Ethereum users are allowed to call a smart contract's functions, and what state the contract has to be in, before defining it. 
  \item {\bf{Assets:}}  Flint supports special operations for handling Assets such as Wei in smart contracts. Transfer operations are performed atomically, and ensure that the state of a contract is always consistent. In particular, Assets in Flint cannot be accidentally created, duplicated, or destroyed, but they can be atomically split, merged, and transferred to other Asset variables. Using Asset types avoids a class of vulnerabilities in which smart contracts' internal state does not accurately represent their true Wei balance. 
  
   \item{\bf{Wei is an asset:}} In Solidity, Wei values are represented as integers rather than a dedicated type, allowing accidental conversions between numbers and currency. This can lead to inconsistent states, in which the actual balance of the smart contract is incorrect.  
   
   \item{\bf{Static typing:}} Given that contracts cannot be corrected, type errors need to be found before contracts are released.
  
  \item{\bf{Modifiers:}} Flint's code is by default private and immutable. A programmer has to explicitly override either of these defaults. It is a compiler error to declare something mutable that isn't changed by the contract.
  
  \item{\bf{Safe Arithmetic:}} Integer overflow causes an exception and contract execution to terminate. There are also cyclic versions of the operators, but a programmer would have to use these special operators explicitly.
  
  \item{\bf{Loops are finite:}} The only loop construct is a for-in loop which is used to iterate over arrays, dictionaries and ranges.
  
  \item{\bf{Initialisers:}} Contracts and structs must define public initialisers, and all state properties will be initialised during their execution.
  
  \item{\bf{Limited Fallback Functions:}} Fallback functions cannot change any state. Default fallback functions rollback the contract.
   
 \end{enumerate}

As recommended by the Ethereum Foundation, we implemented a compiler for Flint which produces EVM bytecode via Solidity's intermediate representation Yul\cite{iulia,yul}. To fit into the existing Ethereum ecosystem, we use the Solidity Application Binary Interface (ABI) and leverage Ethereum's existing cryptographic schemes to use Ethereum user addresses to protect from rogue callers.  Our novel protection system enables static checks on internal calls and runtime checks on external calls.

To evaluate Flint, we translated existing smart contracts and showed the resulting code to be more concise. To assess safety, we ran analysis tools on the bytecode produced and show that a certain class of vulnerabilities cannot be reproduced in Flint. We also assessed the performance of our main safety features.

\section{Solidity: Current State of Play}

Solidity \cite{soliditydocumentation} is statically-typed and imperative. With syntax inspired by JavaScript, Solidity provides a rich set of constructs and it is this expressivity that is visible in many of the bugs.  Avoiding the vulnerabilities that have been exposed by the flawed contracts is critical in the design of Flint.
	
A Solidity contract is similar to an object-oriented class, which can inherit functionality from other classes. Solidity provides integers, addresses, fixed-size arrays, dynamic arrays, and dictionaries (\texttt{mapping}). It is also possible for programmers to define their own types (\texttt{struct}) and their own \textit{interfaces}. A contract  contains storage fields, event declarations, and function declarations.

\textit{Function modifiers} such as \texttt{require} can be used to check preconditions before entering a function's body. If they fail, execution of the contract stops, and the sender receives an exception. Modifiers may mutate the contract's state. 

Functions return a specified number of values.  A function's signature can contain attributes to specify how they are allowed to access the contract's state, and their visibility. Functions declared without visibility modifiers are implicitly \texttt{public}.

Contracts can define an unnamed \emph{fallback function}. When an account calls  a function of a contract which has not been defined, the fallback function is called. Calls can be delegated in which case mutation of state in the target function call is performed in the \emph{caller}'s storage. So programmers who mistype a function's name may unintentionally end up updating state in the wrong storage space.
 
\subsection{Attacks Against Solidity Contracts}\label{solidityattacks}

The design of the Solidity language itself has contributed to it being vulnerable to attack.\cite{dao_wired}   \textit{``Solidity was introducing security flaws into contracts that were not only missed by the community, but missed by the designers of the language themselves.''}\cite{philip_daian} One problem is unintuitive semantics~\cite{philip_daian_blog} of the \texttt{call} method: \textit{``you cannot assume anything about the state of your contract after the external call is executed.''}  We highlight problems with example vulnerable contracts\cite{atzei2017survey}:

\begin{enumerate}

\item{\bf{Call Re-entrancy:}} {\textit{TheDAO Attack.}}\label{thedao}
The attack on TheDAO was possible because Solidity does not have a way to make  transfer of Ether and internal bookkeeping atomic, and  because function calls are synchronous --  thus a function which calls another contract may be re-entered before terminating. In its {\tt refund} function, TheDAO contract sends Ether to a client, and then does the  corresponding internal bookkeeping.  A malicious client re-enters {\tt refund} as soon as it receives the Ether, and thus Ether is repeatedly sent, the bookkeeping is not performed, and the process is repeated until {\tt refund} runs out of gas or the contract runs out of Ether.

\item{\bf{Visibility Modifiers: }}{\textit{the First Parity Multi-sig Wallet Hack.}}\label{paritymultisig1}
At the time of the attack, Solidity initialisers could be called anywhere any number of times.  The confusing semantics of delegation and fallback functions means that external contracts can update state not intended by the contract developer.
Parity accounts control a common wallet. An attacker exploited the wallet to steal over \$80 million~\cite{parity1}. The library code Parity provided was written as a contract and users creating their own wallet contract, could delegate all the functions to the library instance. There is an initialising function to set the owner of a wallet that should only be called in the constructor that sets up a new wallet.  Unfortunately, due to the semantics of delegation and fallback functions, the caller can call an initialiser at any time and set themselves as the owner of the entire Parity contract.

\item{\bf{Contract as a Library: }}{\textit{the Second Parity Multi-sig Wallet Hack.}} At the time of the attack, Solidity libraries were just contracts so they could have updateable state. A second problem was that initialisers do not have to be called. The previous Parity Multi-sig contract was affected by another attack, which caused the loss of approximately \$260 million. The initialisation function now did have a modifier (\texttt{only\_uninitialised}). The assumption from the developers was that their \texttt{WalletLibrary} would only be interacted with through delegate calls (mutating the caller's state rather than the library's state). However, \texttt{WalletLibrary}, which was just an ordinary contract, didn't actually call the initialisation function, so an attacker was able to call the initialiser, set the owner to itself and then terminate the contract. All of the other wallets delegating their calls to \texttt{WalletLibrary} were frozen: the instance of \texttt{WalletLibrary} they were delegating was destroyed. There were two problems: the first was that an initialiser did not have to be called at all, and the second was that a user was able to modify the state of a smart contract which was only meant to be used as a stateless library. 

\item{\bf{Unchecked Calls: }}{\textit{King of the Ether Throne.}}  Solidity does not require return values to be checked. This contract keeps track of a current king. An account needs to pay more than the king paid in order to dethrone him and when a king is dethroned, he gets sent his stake back. This will not happen if the contract runs out of gas and the boolean indicating whether the transaction was successful is not checked. When an out of gas exception occurs, the Ether is returned to the caller and as the king is not the caller he doesn't get his Ether back.

\item{\bf{Arithmetic Overflows: }}{\textit{Proof of Weak Hands Coin.}}\label{powhc}
Arithmetic operators have wrap-around semantics. This contract implements a currency. About \$476K was lost due to an arithmetic overflow following an addition. 
\end{enumerate}

The attacks can mostly be attributed to a mix of human error and unintuitive language semantics.  Understanding the pitfalls that some Solidity programs have fallen into, was the starting point of the design of Flint. We now discuss how Flint makes the development of vulnerable contracts harder to write accidentally.

\begin{enumerate}

\item{\bf{Call Reentrancy:}} {\textit{TheDAO Attack.}}
Flint uses an @payable function annotation, for currency transfer. @payable requires a Wei value rather than an integer. Wei is transferred atomically so there could not be a transfer before updating the contract's bookkeeping. Also there is limited re-entrancy, and our fallback functions cannot update state.
\item{\bf{Visibility Modifiers: }}{\textit{the First Parity Multi-sig Wallet Hack.}}
Flint requires initialisers which have to initialise all state and which get called exactly once and that is before use. A caller could overwrite these values, if the contract developer made them visible, but the default is everything is private and immutable.
\item{\bf{Contract as a Library: }}{\textit{the Second Parity Multi-sig Wallet Hack.}} Libraries are stateless, Flint's default fallback functions just rollback, and Flint enforces the initialisation of contract state properties inside the init call which has to occur on the instantiation of contracts and structs.
\item{\bf{Unchecked Calls: }}{\textit{King of the Ether Throne.}} In Flint, the King of the Ether Throne vulnerability caused by discarding the result of {\tt{send}} could not happen as it is a compile error to discard the result of a function call. There is also a possible denial of service attack\cite{atzei2017survey} and the corresponding Flint program would only suffer from this problem, if the non re-entrancy of external calls were explicitly changed to allow re-entrancy. 
\item{\bf{Arithmetic Overflows: }}{\textit{Proof of Weak Hands Coin.}}
Flint's default integers do not overflow. So a Flint programmer could cause an arithmetic overflow, but they have to do it explicitly by using the alternative integers operators.
\end{enumerate}

The developers of Solidity apparently shared some of our concerns but have been proposing very different solutions to the problems. For example, having a modifier (\texttt{only\_uninitialised}) although sufficient to prevent a constructor from being called twice, is not sufficient to get it called once, whereas Flint's constructor is called exactly once and at the appropriate time. Solidity's new inclusion of some SMT verification of arithmetic\cite{smt-solidity} should catch overflows, but Flint's philosophy is to use safe arithmetic in the first place. We do not see how the Solidity developers are going to be able to alter their language to prevent the confusing semantics of fallbacks and delegation. Both languages have assertions. Solidity programmers need to insert them in their code in the places that Flint programmers will use safe arithmetic, protection blocks and Assets.

\section{Flint by Example}\label{flintproject}

We present Flint, influenced by Swift\cite{swift} syntax, for writing safe Ethereum smart contracts. 
Like in Solidity, a Flint smart contract's state is represented by its fields, or \textit{state properties}.
Unlike in Solidity, state properties are declared in isolation from functions so that programmers can easily ensure that no unnecessary state properties are declared. Like in Solidity, they are stored in the smart contract's persistent memory (\textit{storage}), with high access costs.

Like in Solidity, a Flint contract's behaviour is characterised by its functions. Unlike Solidity, they are declared within  \textit{protection blocks} rather than at the top level of the contract. This forces programmers to first think about what state the contract needs to be in and which parties should be able to make function calls before defining functions. Functions are by default non-mutating, but can explicitly mutate the contract's state if declared as \texttt{mutating}. Functions are by default private, but those with a \texttt{public} modifier can be called by external Ethereum users. 
The standard library offers an \texttt{Asset trait}, which provides safe atomic operations to handle currency, ensuring the state of smart contracts is always consistent. 

\subsection{Declaring Contracts}~\label{simpledao}

To introduce some features of Flint\footnote{For the full description of Flint see~\cite{guide}.} we start the development of a \texttt{SimpleDAO} contract.
When declaring the contract, we observe how Flint's syntax requires programmers to write their smart contact in a specific sequence of steps.

\begin{figure}
\includegraphics[width=100mm]{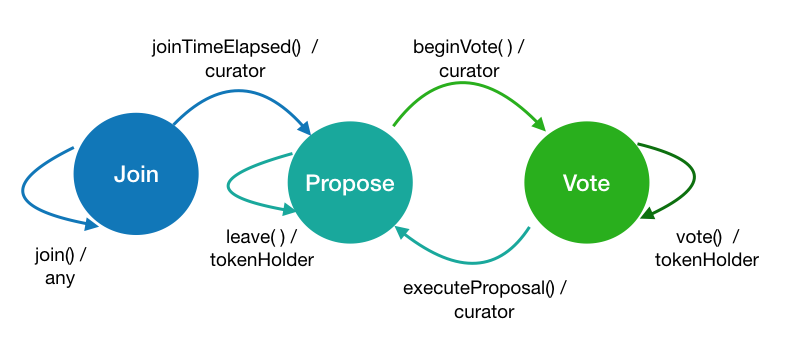}
    \centering
  \caption{DAO State Changes}
  \label{compilerstages}
\end{figure} 

\begin{description} 
\item[1. Declaring the contract's possible states.] 

The contract needs to enable users to join ({\tt{Join}}), if they have joined to be able to propose transfers of {\tt{Wei}} ({\tt{Propose}}), and if a transfer has been proposed to vote ({\tt{Vote}}). 
This \textit{typestate} appears in the contract header. The \texttt{Address} type represents an Ethereum address (a user or another contract).
\begin{lstlisting}[language=Flint, name=SimpleDAO]
contract SimpleDAO (Join, Propose, Vote) {
  var curator: Address // a very simple consensus mechanism
  visible var proposal: Int = 0
  var proposals: [Proposal] = [] // a list of all proposals to transfer Wei
  var balances: [Address: Wei] = [:] // rmembers' Wei balances
}
\end{lstlisting}

\item [2. Declaring the protection blocks.] 
Functions of a contract are declared within protection blocks, which restrict when the enclosed functions are allowed to be called. There are two elements to protection blocks: the caller protection and the optional typestate protection. A protection block declaration has to include the contract name (\texttt{SimpleDAO}) followed by {\tt{@(typestate)}}(e.g., \texttt{Join}) , followed by a \texttt{::} and admisable callers (e.g., \texttt{curator}).

The first protection block for this contract is for setting up a new SimpleDao contract.  Anyone ({\texttt{any}}) may do this and the caller's address is bound to the \texttt{caller} local variable. This initialiser can only be called when the contract is created as all other contracts have a typestate parameter associated with them. The second protection block is for joining an existing contract. The third is for the \texttt{curator} to explicitly change the state from {\texttt{Join}} to {\texttt{Propose}}, closing off the SimpleDAO from accepting new members. The final two protection blocks are for processing proposals and votes respectively. The bodies of the protection blocks are completed later.

\begin{lstlisting}[language=Flint, name=SimpleDAO]
SimpleDAO @(any) :: caller <- (any) { ... }
SimpleDAO @(Join) :: caller <- (any) { ... }
SimpleDAO @(Join) :: (curator) { ...}
SimpleDAO @(Propose) :: caller <- (tokenHolder) { ... }
SimpleDAO @(Propose) :: caller <- (curator) { ... }
SimpleDAO @(Vote) :: caller <- (tokenHolder) { ... }
 \end{lstlisting}

\item[Declaring the global structs.] Struct values can be declared as state properties or local variables, and are initialised through their initialiser. When stored as a state property, the struct's data resides in EVM storage. When stored as a local variable, it resides in EVM memory, and a pointer is allocated on the EVM stack. A struct's functions  are not explicitly protected by being in protection blocks, rather they will be protected by the contract functions that call them. Our contract only needs a {\texttt{Proposal}} struct:

\begin{lstlisting}[language=Flint, name=SimpleDAO]
struct Proposal {
  var proposer: Address
  var payout: Int
  var recipient: Address
  var yea: Int = 0
  var nay: Int = 0
  var finished: Bool = false
  var success: Bool = false
  var voted: [Address: Bool] = [:]

  mutating init(proposer: Address, payout: Int, recipient: Address) { self.proposer = proposer
    self.payout = payout
    self.recipient = recipient
  }
}
\end{lstlisting}

\item [3. Declaring the functions.]

Finally, functions are declared within the protection blocks. For example, the third protection block where the {\tt{curator}} stops taking new members is only one function. The complete code appears in Appendix~\ref{simple}.

\begin{lstlisting}[language=Flint, name=SimpleDAOfunc]
SimpleDAO @(Join) :: (curator) {
   public mutating func joinTimeElapsed() {
     become Propose
   }
 }
 \end{lstlisting}
\end{description}

\subsection{Additional Language Features}

\paragraph{\bf{Initialisation}\label{Initialisation}}
Each smart contract and struct must define exactly one public initialiser. All of the state properties must be initialised before the initialiser returns. State properties can be declared with a default value and constants must be assigned exactly once. 

\paragraph{\bf{Type System}}
Flint is a statically-typed language, with no support for sub-typing.  
Flint supports basic types and dynamic types (Figure \ref{types}). Dynamic types can be passed by value or by reference ({\verb+&+}).

\begin{figure}
\centering
\begin{tabular}{ | p{6em} | p{30em} | }	
	\hline
Type & Description\\ \hline \hline
\texttt{Address} & 160-bit Ethereum address\\ \hline
\texttt{Int}  & 256-bit unsigned integer\\ \hline
\texttt{Bool}  & Boolean value\\ \hline
\texttt{String}  & String value\\ \hline
\texttt{Void}  & Void value\\ \hline
Fixed-size Array & Fixed-size memory block containing elements of the same type. \texttt{T[n]} refers to an array of size \texttt{n}, of element type \texttt{T}. \\ \hline
Array & Dynamically-sized array. \texttt{[T]} refers to an array of element type \texttt{T}. \\ \hline
Dictionary & Dynamically-size mappings from one key type to a value type. \texttt{[K: V]} is a dictionary of key type \texttt{K} and value type \texttt{V}.\\ \hline
Structs & Struct values, including \texttt{Wei}, are considered to be of dynamic type.\\ \hline
\end{tabular}
\caption{Flint Types}
\label{types}
\end{figure}

Flint has traits which are based on Rust\cite{rust} traits. There are both contract and struct traits and they describe the partial behaviour of the contracts and structs which conform to them. Contracts and structs can conform to multiple traits as long as there is at most one function body for any given function. An example {\tt{struct trait}} with a conforming {\tt{struct}} is given below. A contract trait would be similar with the keyword {\tt{contract}} replacing {\tt{struct}}.

\begin{lstlisting}[language=Flint, label=strait]
struct trait Animal {
  // Must have an empty and named initialiser
  public init()
  public init(name: String)
  func isNamed() -> Bool
  public func name() -> String
  public func noise() -> String
  public func speak() -> String {
    if isNamed() {return name()} else {return noise()}
  }
}

struct Person: Animal {
  let name: String
  public init()  self.name = "John Doe"}
  public init(name: String) {self.name = name}
  func isNamed() -> Bool {return true}
  public func name() -> String {return self.name}
  public func noise() -> String {return "Huh?"}
  // Person can also have functions in addition to Animal
  public func greet() -> String {return "Hi"}
}
\end{lstlisting}

\begin{figure}
\centering
\begin{tabular}{ | p{12em} | p{25em} | }
	\hline
Function & Description\\ \hline \hline
\texttt{send(address: Address, value: inout Wei)} & Sends \texttt{value} Wei to the Ethereum address \texttt{address}, and clears the contents of \texttt{value}. \\ \hline
\texttt{fatalError()} & Terminates the transactions with an exception, and revert any state changes. \\ \hline
\texttt{assert(condition: Bool)} & Ensures \texttt{condition} holds, cause a \texttt{fatalError()}. \\ \hline
\end{tabular}
\caption{Flint Global Functions}
\label{globalfunctions}
\end{figure}

\paragraph{\bf{The Standard Library}}

We also define three global functions, shown in Figure~\ref{globalfunctions}. Global functions are defined in the special \texttt{Flint\$Global} struct in \texttt{stdlib/Global.flint} and are imported globally by the compiler. There is an {\tt{Asset}} trait defined in the standard library (see Appendix~\ref{assets}) as well as {\tt{Wei}} which conforms to it. Compiler checks ensure that contracts use {\tt{Wei}} from the Standard Library.
\subsubsection{Asset Traits}
\begin{description}
  \item [No Unprivileged Creation] It is not possible to create an asset of non-zero quantity without transferring it from another asset.
  \item [No Unprivileged Destruction] It is not possible to decrease the quantity of an asset without transferring it to another asset.
  \item [Safe Internal Transfers] Transferring a quantity of an asset from one variable to another within the same smart contract does not change the smart contract's total quantity of the asset.
  \item [Safe External Transfers] Transferring a quantity $q$ of an asset $A$ from a smart contract $S$ to an external Ethereum address decreases $S$'s representation of the total quantity of $A$ by $q$. Sending a quantity $q'$ of an asset $A$ to $S$ increases $S$'s representation of the total quantity of $A$ by $q'$.
\end{description}

\paragraph{\bf{Safe Arithmetic Operators}}

The \texttt{+}, \texttt{-}, and \texttt{*} operators throw an exception and abort execution of the smart contract when an overflow occurs. The \texttt{/} operator implements integer division. For the rare cases where the intended behaviour is cyclic, Flint also supports wrap-around operators, \texttt{\&+}, \texttt{\&-}, and \texttt{\&*}.

\paragraph{\bf{@payable Annotation}}

Similar to what Solidity does, when a user creates a transaction to call a function, Ether can be sent by using \texttt{@payable}. A single parameter marked \texttt{implicit} of type \texttt{Wei} must be declared; \texttt{implicit} parameters expose information from the Ethereum transaction to the developer of the smart contract, without using globally accessible variables, such as \texttt{msg.value} in Solidity. 
\forget{In Listing~ref{payable}, the number of {\tt{Wei}} attached to the Ethereum transaction performing the \texttt{receiveMoney} call is bound to the implicit variable \texttt{value}.}
\begin{lstlisting}[language=Flint, label=payable]
@payable
public func receiveMoney(implicit value: Wei) {
  doSomething(value)
}
\end{lstlisting}

\paragraph{\bf{Events}}

JavaScript applications can listen to events emitted by an Ethereum
smart contract. Flint provides the same functionality with slightly different syntax from that provided by Solidity.

\begin{lstlisting}[language=Flint, label=events]
contract Bank {
  var balances: [Address: Int]
  let didCompleteTransfer: Event<Address, Address, Int> // (origin, destination, amount)
}
Bank :: caller <- (any) {
  mutating func transfer(destination: Address, amount: Int) {
    // Omitting the code which performs the transfer.
    emit didCompleteTransfer(caller, destination, amount)
  }
}
\end{lstlisting}

\section{Implementation}

We implemented \texttt{flintc}, a compiler for Flint which runs on Linux and macOS. The compiler's source code (about 25,000 lines of Swift\cite{swift} code) is open source and available on GitHub\cite{flintgithub}. 

The compiler stages are illustrated in Figure \ref{compilerstages}. Programs are analysed, compiled to the Yul intermediate representation, and finally to EVM bytecode. By embedding Yul in a Solidity file, tools built for Solidity work with Flint. We compile Yul code using the Solc\cite{soliditydocumentation} compiler. The tokeniser and parser are hand written as this has enabled us to provide better error messages.

\begin{figure}
\includegraphics[width=100mm,scale= 1]{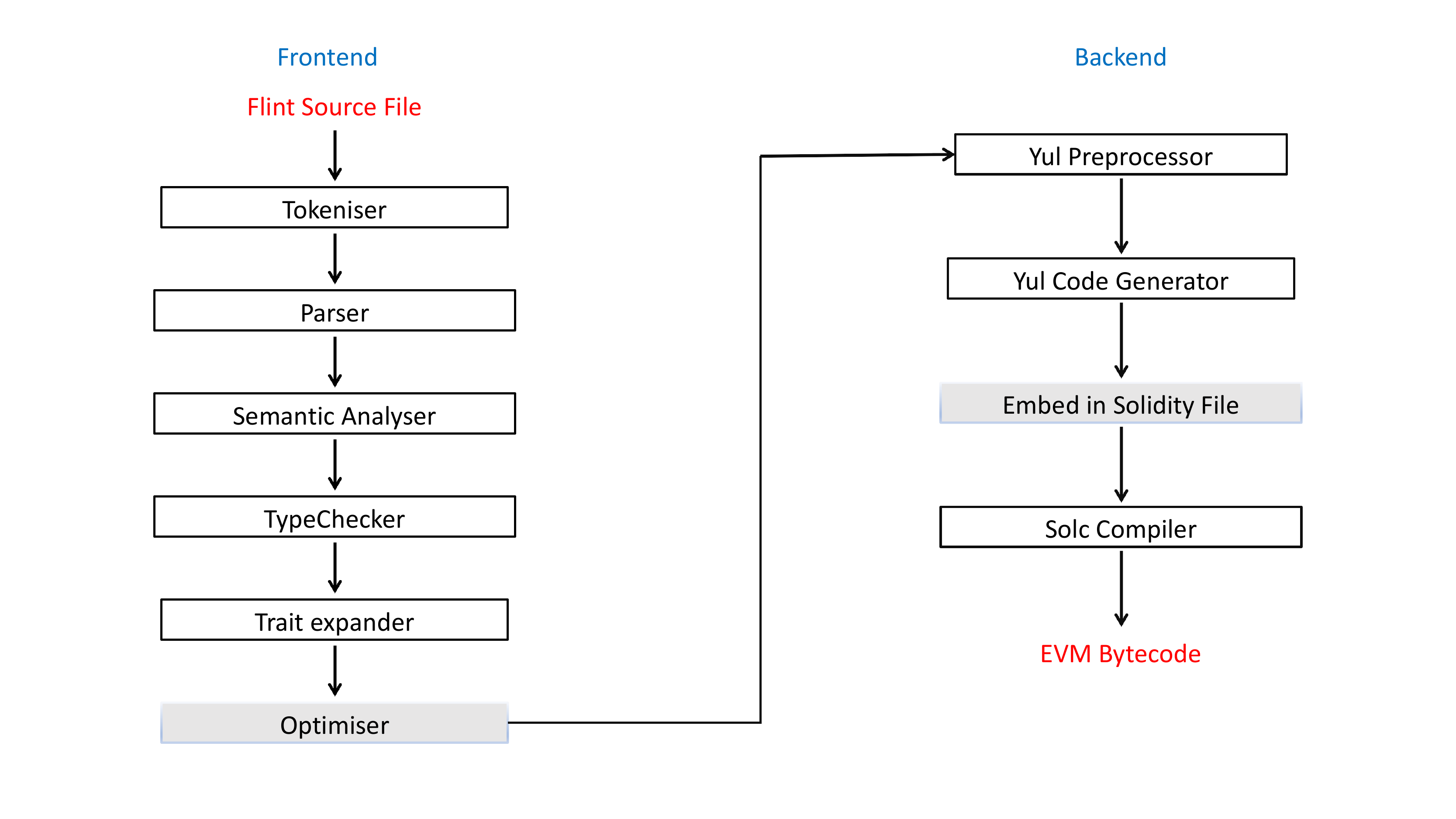}
    \centering
  \caption{Compiler Stages}
  \label{compilerstages}
\end{figure} 
The \textit{Parser} has an additional step at the end which populates an environment. The environment contains information about the discovered types in the compilation step. Contract/Structure types contain information about the conforming traits, functions and fields. Contracts also contain additional information about any declared typestates. This provides sufficient information to AST Passes for the checking of non-trivial program properties.

Our code architecture decouples AST traversal and node processing. The traditional visitor pattern\cite{palsberg1998essence} leverages dynamic method dispatch mechanisms to separate node processing logic from tree traversal logic. However, visited nodes invoke the visitor on their children nodes manually. Our nodes do not have references to visitors, rather we have an architecture which uses a single \textit{AST Visitor}, and multiple \textit{AST Passes}. Each AST Pass implements a process function for each type of AST node. The AST Visitor updates the tree using the nodes returned by the AST Pass, collects diagnostics, and propagates contextual information. The  AST passes are Semantic Analyser, Type Checker, TraitResolver, and Yul Preprocessor.

The Semantic Analysis AST Pass verifies the correctness of the input program. This performs the static checks for caller and typestate protection, traits,  \texttt{mutating} functions, verifying whether there are uses of undefined variables, etc. 
Appendix~\ref{diagnostics} contains some examples of errors and warning diagnostics that are produced during this pass.

As there is no subtyping or polymorphism, the Type Checking pass is straightforward. 
Appendix~\ref{diagnostics} also contains examples of type error messages.

\subsection{Code Generation and Runtime}\label{implementation}

Before the code generation phase, like in most object oriented language compilers, we apply a preprocessing step which mangles function names (needed for disambiguating for overloading and for functions coming from different contracts or structs). For each parameter which can be passed by value or by reference an \texttt{isMem} parameter is added.
At this stage traits are embedded in conforming structures and contracts.

We implement a code generating function per AST Node, which takes an AST node as a parameter, and returns its Yul representation. The \texttt{YulStruct} function generates code for a Flint struct. Similarly, we implement \texttt{YulFunction}, \texttt{YulAssignment}, \texttt{YulExpression}, etc.

\subsection{Functions and Application Binary Interface}\label{abi}

Flint's Application Binary Interface (ABI) specifies at the bytecode level how Ethereum users and other smart contracts can call the public functions of a Flint smart contract. Flint follows Solidity's ABI, therefore Flint and Solidity contracts can interoperate. Users can call a smart contract's function on Ethereum by creating a transaction, and specifying which function to call with which arguments in the transaction payload. Transaction payloads are raw bytes, thus the data needs to be encoded.
Specifying which function to call is done via encoding the function's signature. Function arguments are appended to the function signature hash as a hexadecimal value. Calling $f$ with arguments \texttt{100} and \texttt{true} would be encoded as \texttt{0x64} and \texttt{1}, padded with zeros to fill a 256-bit value. An Ethereum user or another smart contract can thus call $f$ with arguments $100$ and $true$ by entering the following value in the transaction payload (without newline characters):

\begin{lstlisting}[style=bash]
0x13d1aa2e
0000000000000000000000000000000000000000000000000000000000000064
0000000000000000000000000000000000000000000000000000000000000001
\end{lstlisting}

A struct value passed as an \texttt{inout} (reference) argument to a function is an implicit reference to either an EVM memory location or an EVM storage location. When accessing the memory location, the runtime needs to know whether it should read the value from memory or from storage. To support this, when a struct is passed by reference to a function, an extra boolean argument, specifying the location of the reference, is inserted in the argument list of the function call. 
The storage and memory of a Flint smart contract are organised similarly. The runtime functions \texttt{load} and \texttt{store} work for both variants.
A contract's state properties are stored in EVM storage sequentially, except for values in dynamic arrays and dictionaries. 
Local variables are stored in memory, and are allocated dynamically. The \texttt{allocateMemory} runtime function reserves a number of bytes in memory, and returns the start pointer of the block. The first 64 bytes of memory (8 words) are reserved as scratch space and can be used to perform temporary computation, or load values into memory to compute \texttt{sha3} hashes or emit Ethereum events. Memory location \texttt{0x40} (64th byte) holds a pointer to the next available memory location (initially \texttt{0x60}). 

State properties of smart contracts are stored contiguously in storage, starting at location 0. Each state property occupies one word (32 bytes) in the case of basic types, or multiple words when storing structs or fixed-sized arrays. Structs can also be stored in memory. Dynamically-sized types, such as arrays and dictionaries, are not necessarily allocated contiguously.  Storage accesses yield the same gas cost regardless of which location is accessed. 

The Flint runtime contains 20 runtime functions to perform low-level operations. A runtime function $f$ can be called from the Flint standard library using \texttt{flint\$f}, but not from user-defined code. 

Protection checks (both callers and typestate) are performed at compile-time for internal (same contract) function calls (except {\tt{try}} code) while the protection of foreign contracts calling into Flint contracts, (and  {\tt{try}} code) are checked at runtime. To enable the verification, runtime checks are inserted immediately before running the function's body. If the caller's address is not in the set of caller protections required to call the function, an exception is thrown and the call is aborted (the \texttt{REVERT} opcode is executed).

Typestate protection checks are performed similarly to caller protection checks and are done at compile-time for internal functions and at runtime for foreign contracts calling into Flint contracts. At compile time an internal typestate enumeration is generated denoting all of the user defined states and the internal states. An internal contract state property is defined by the compiler with that type. When a runtime check needs to be performed the compiler will generate a comparison against the set of valid typestates, and if it is not valid, an exception is thrown and the call is aborted (the \texttt{REVERT} opcode is executed).

Typestate can only be changed in user-defined code through {\tt{become StateIdentifier}} statements. The compiler will generate code to update the contract state property to the appropriate  enumeration value.

\subsection{Intermediate Representation Organisation}
To generate the IR code it is done in the following order:
\begin{enumerate}
\item  {\emph{Contract function definitions.}} Code is generated for user-defined contract functions, one IR function for each Flint contract function. The Yul code does not include explicit declarations of state properties, as accesses to storage properties in functions are represented as static offsets into memory.
\item{\emph{Struct function definitions.}} The function code for each user-defined and standard library struct is included next. 
\item {\emph{Runtime functions.}} Finally, we include the definition of any runtime functions.
\end{enumerate}

\forget{
\begin{figure}
  \centering
  \includegraphics[width=320px]{language/flintCI}
  \caption{Overview of Flint's Continuous Integration infrastructure}
  \label{flintci}
\end{figure}

\subsection{Toolchain}\label{compilertesting}

\forget{Our development workflow is described in Figure~\ref{flintci}.} 
To help with the correctness of our implementation, we create an automated test suite for the different stages of our compiler and the produced bytecode. We also implement an automated deployment infrastructure for \texttt{flintc} binaries using GitHub Releases.

 We use Travis CI\cite{travisci} to run our tests. To test the compiler, we implement tests verifying whether the AST produced by the parser is correct, and tests to verify the compiler produces the valid warnings and errors. In addition, we test the behaviour of the bytecode produced by the Flint compiler by running a local simulated Ethereum blockchain.

We use the Lite\cite{swiftlite} library to automatically find and run tests. We use the Truffle\cite{truffle} library for Solidity smart contracts to test the behaviour of the bytecode the Flint compiler produces. We embed our generated IR code in a Solidity file, and use the generated Solidity interface for Truffle to interpret Flint contracts as Solidity contracts. We write our tests in JavaScript using the Web3\cite{web3js} library to call smart contract functions from Truffle. 
}

\section{Evaluation}

In this section we compare Flint to Solidity.  We were able to use the same analysers Oyente\cite{makingsmartcontractssmarter} and Mythril\cite{Mythril}  by embedding Flint's IR code in a Solidity file. The Solidity and Flint gas costs have been retrieved from executing the calls on our simulated Ethereum network.

The most important features of Flint for writing robust code are protection blocks (with callers and typestate), {\tt{Assets}}, and safe arithmetic so these are the features we examine in some detail. We provide small example programs for each construct in both languages so that we can compare code styles, potential for bugs as discovered by Solidity dynamic analysis tools, and gas usage. Solidity has run-time type checking which should add a performance overhead, whereas Flint currently has no optimisation and there is a performance overhead caused by embedding our Yul code in a Solidity file. 

\subsection{Caller Protections}\label{callercapseval}

To compare caller protections we define simple functions some of which can be called by any user, some by \texttt{owner}, and some by any customer in the \texttt{customers} array. In Solidity, we implement two modifiers, \texttt{onlyOwner} and \texttt{anyCustomer}, which check whether the caller is \texttt{owner} or is in the \texttt{customers} array, respectively.

\textsc{Solidity}
\begin{lstlisting}[language=Solidity, label={ownercaps}]
modifier onlyOwner {
  require(msg.sender == owner);
  _;
}
modifier anyCustomer {
  uint numCustomers = customers.length;
  bool found = false;
  for (uint i = 0; i < numCustomers; i++) {
    if (customers[i] == msg.sender) { found = true; }
  }
  require(found);
  _;
}
contract Callers {
  address owner;
  address[] customers;
  uint256 counter;
  modifier anyOwner {...}
  modifier anyCustomer {...}
  function anyUser() public constant returns(uint256) {...}
  function ownerCall(uint counter) public constant onlyOwner returns(uint256) {...}
  function customerCall(uint counter) public constant anyCustomer returns(uint256) {...}
  function ownerInternalCalls() public onlyOwner {...}
  function customerInternalCalls() public anyCustomer {...}
}
\end{lstlisting}

The organisation of the smart contracts presents notable differences. In the Solidity code (above), the state and the functions are all defined at the top level of the contract. Solidity does not enforce the inclusion of the user-defined modifiers we have specified in some of the function signatures. In the more concise Flint code (below) the contract itself and the protection blocks are at the top level. 

\textsc{Flint}
\begin{lstlisting}[language=Flint]
contract Callers {
  var owner: Address
  var customers: [Address]
  var numCustomers: Int
  var counter: Int
}
Callers :: owner <- (any) {
  mutating public func addCustomer(customer: Address) {...}
  public func anyUser() -> Int {...}
}
Callers :: (owner) {
  public func ownerCall(counter: Int) -> Int {...}
  public mutating func ownerInternalCalls() {...}
}
Callers :: (customers) {
  public func customerCall(counter: Int) -> Int {...}
  public mutating func customerInternalCalls() {...}
}
\end{lstlisting}
In Figure~\ref{callercapsperformance} we compare the gas costs of executing each function in the Solidity and Flint contracts.  Flint is significantly faster when functions perform internal calls to functions which require the same caller protection. This is expected, as the caller protection check is only executed once.

The gas costs are similar for simple examples using a caller protection backed by a single address. When calling a function which calls multiple functions, requiring the same caller protection, Flint is up to 12 times faster than Solidity.

For caller protections backed by an array, the results aren't as good. Flint code is cheaper to run when the array is relatively small, but becomes more expensive when it is large. In the example which performs a large number of internal calls, the Flint code is up to two times faster as Flint performs a single runtime check, whereas Solidity performs one per function call.

\begin{figure}
\begin{tabular}{ | p{7em} | p{4em} | p{3em} | p{5em} | p{14em} | }
	\hline
Operation & Solidity  & Flint & Difference (Solidity vs Flint) & Possible Explanation\\ \hline \hline
Deploying & 438121 &  514268 & S: -14.8\% & Flint does not optimise code.\\ \hline
Any call & 268 &  283 & S: -5\% V: -32\% & Determining which function was called is slightly faster in Flint.\\ \hline
Owner call & 633 &  836 & S: -24\%  V: -22\% & \\ \hline
Owner many internal calls & 32662 &  28673 & S: +14\%  V: +496\% & Only one dynamic caller protection check is performed in Flint. \\ \hline
Customer call (5 \texttt{customers}) & 3791 &  1707 & S: +122\%  V: +158\% & Flint's array iteration algorithm is better suited for small arrays.\\ \hline
Customer call (20 \texttt{customers}) & 13136 & 18047 & S: -27\%  V: -37\% & Solidity's array iteration algorithm suits larger arrays.\\ \hline
Customer call (50 \texttt{customers}) & 31434 &  43847 & S: -28\%  V: -43\% & Same as above.\\ \hline
Customer many calls (5 \texttt{customers}) & 168804 &  54426 & S: +210\%  V: +1068\% & Only one dynamic caller protection check is performed in Flint.\\ \hline
Customer many calls (20 \texttt{customers}) & 510129 &  67326 & S: +658\%  V: +2327\% & Same as above.\\ \hline
Customer many calls (50 \texttt{customers}) & 1192779 &  93126 & S: +1181\%  V: +3792\% & Same as above.\\ \hline
\end{tabular}
\caption{Gas Costs for \texttt{Callers}.}
\label{callercapsperformance}
\end{figure}

We compare the result of the analyses performed by the dynamic Solidity analysis tools Oyente and Mythril.  Both analysis tools detect if the \texttt{customers} array has the potential to become too large. While in Solidity the array will silently overflow, in Flint it would throw an exception.

\subsection{Typestate Protection}\label{typestateseval}

To examine typestate we define simple functions which can be called in only one of two states, as well as simple functions to transition between the two states. In Solidity, we implement a modifier, \texttt{atStage} which checks that the contract is in the correct State.

\textsc{Solidity}
\begin{lstlisting}[language=Solidity, label={ownercaps}]
modifier atStage(Stages _stage) {
	require(
		stage == _stage,
		"Function cannot be called at this time."
	);
	_;
}
contract StateMachine {
    enum Stages {
        Stage1,
        Stage2
    }
    Stages public stage = Stages.Stage1;
    modifier atStage(Stages _stage) { ... }
    function nextStage() public { ... }
    function Stage1Function() public atStage(Stages.Stage1) { ... }
    function Stage1ComplexFunction() public atStage(Stages.Stage1) { ... }
    function Stage2Function() public atStage(Stages.Stage2) { ... }
}
\end{lstlisting}

\textsc{Flint}
\begin{lstlisting}[language=Flint]
contract States (State1, State2) { ... }
States @(any) :: (any) {
	public init() {
		become State1;
	}
}
States @(State1) :: (any) {
	public mutating func State2Transition()  { ... }
	public func State1Function() { ... }
	public func State1ComplexFunction() { ... }
}
States @(State2) :: (any) {
	public mutating func State1Transition() { ... }
	public func State2Function() { ... }
}
\end{lstlisting}

We compare the gas costs of executing each function in the Solidity and Flint contracts in Figure~\ref{typestatesperformance}. For simple external calls requiring a state check, the gas consumption for both Flint and Solidity is low, although Flint has a slightly higher consumption, probably due to function selection not being optimised. State transition takes less gas in Flint as only a single memory store has to take place while the compiled Solidity code makes multiple changes.

For operations involving many internal calls, Flint has a much lower gas consumption, with the Flint code consuming only about half as much as the comparable Solidity code. This lower overhead is because Flint only has to perform a single runtime check at the start of the transaction rather than many checks for each function as the code progresses.

Flint's source code is more concise, whereas the compiled code is slightly larger and deployment therefore takes more gas.

\begin{figure}
\begin{tabular}{ | p{7em} | p{4em} | p{3em} | p{5em} | p{14em} | }
	\hline
Operation & Solidity  & Flint & Difference (Solidity vs Flint) & Possible Explanation\\ \hline \hline
Deploying & 239350 &  282031 & S: -15.1\% & Flint does not optimise code.\\ \hline
External call & 650 &  970 & S: -33.0\% & Flint function selector less optimised\\ \hline
Many internal calls (50) & 18434 & 8729 & S: +111\% & Only one dynamic caller protection check is performed in Flint. \\ \hline
State transition (average) & 20855 &  6696 & S: +211\% & Fewer state changes performed in Flint\\ \hline
\end{tabular}
\caption{Gas Costs for \texttt{Typestates}.}
\label{typestatesperformance}
\end{figure}

\subsection{{\tt{Asset}} Types and Safe Arithmetic Operations}\label{bank}

In this section we are interested in the security of Solidity and Flint contracts when handling Wei, and measure whether Flint {\tt{Asset}}s introduce performance penalties. We define the \texttt{Bank} smart contract where customers can send {\tt{Wei} }to the contract, and \texttt{Bank} keeps track of how much each customer has sent. Customers can then withdraw their {\tt{Wei}}, or transfer it to another {\tt{Bank}} account. The complete Flint code for {\tt{Bank}} is in Appendix~\ref{AppendixBank}.

The \texttt{deposit} function of both smart contracts are similar, and record received {\tt{Wei}} to state. Flint uses the {\texttt{Wei}} type, whereas as there is no built-in {\tt{Wei}} type in Solidity, currency is held as an integer. It is easy to accidentally add currency to an account, or forget to do so. Using Flint's {\tt{Wei}} {\tt{Asset}}s allows safer and more concise code for transferring funds.

\textsc{Solidity}
\begin{lstlisting}[language=Solidity]
function deposit() anyCustomer public payable {
  balances[msg.sender] += msg.value;
}
\end{lstlisting}

\textsc{Flint}
\begin{lstlisting}[language=Flint]
@payable
public mutating func deposit(implicit value: Wei) {
  balances[account].transfer(&value)
}
\end{lstlisting}

The comparison between \texttt{transfer} functions is similar. An additional advantage of the Flint {\tt{transfer}} is the update of state when the currency is moved is atomic.
 
\textsc{Solidity}
\begin{lstlisting}[language=Solidity]
 function transfer(uint amount, address destination) anyCustomer public {
  require(balances[msg.sender] >= amount);
  balances[destination] += amount;
  balances[msg.sender] -= amount;
}
\end{lstlisting}

\textsc{Flint}
\begin{lstlisting}[language=Flint]
public mutating func transfer(amount: Int, destination: Address) {
  balances[destination].transfer(&balances[account], amount)
}
\end{lstlisting}

The \texttt{withdraw} function is one line shorter in Flint as we do not need to check whether the account holder has enough funds to perform an operation; an exception is thrown if the result of an arithmetic operation overflows. There are cleaner ways to say that a {\tt{withdraw}} failed, but at least it is safe. However, if a Solidity programmer forgot and there wasn't enough in the account to make the withdrawal there would be a silent overflow and the value of the account's {\tt{balance}} would be wrong.

\textsc{Solidity}
\begin{lstlisting}[language=Solidity]
function withdraw(uint amount) anyCustomer public {
  require(balances[msg.sender] >= amount);
  balances[msg.sender] -= amount;
  msg.sender.transfer(amount);
}
\end{lstlisting}

\textsc{Flint}
\begin{lstlisting}[language=Flint]
public mutating func withdraw(amount: Int) {
  // Transfer some Wei from balances[account] into a local variable.
  let w: Wei = Wei(&balances[account], amount)
  // Send the amount back to the Ethereum user.
  send(account, &w)
}
\end{lstlisting}
\forget{In Flint \texttt{Wei}'s initialiser atomically transfers a subset of another {\tt{Wei}} variable to the receiver.}

\begin{figure}
\begin{tabular}{ | p{8em} | p{4em} | p{3em} | p{4em} | p{15em} | }
	\hline
Operation & Solidity & Flint & Difference & Possible Explanation\\ \hline \hline
Deploying & 422589 & 415901 & +1.6\% & Solidity and Flint produce similarly sized binaries.\\ \hline
Register & 40741 & 61528 & -34\% & Solidity is more efficient at adding elements to an array.\\ \hline
Deposit 10 Wei & 21460 & 24002 & -11\% & This is because of the overhead of Flint's safe Asset operations, which prevent overflows and state inconsistencies. \\ \hline
Transfer 80 Wei & 27200 & 30685 & -11\% & Same as above. \\ \hline
Withdraw 5 Wei & 14301 & 17245 & -17\% & Same as above. \\ \hline
Mint 100 Wei & 23098 & 20867 & +10\% &  Similar results.\\ \hline
\end{tabular}
\caption{Gas Costs for \texttt{Bank}}
\label{assetperformance}
\end{figure}

Figure~\ref{assetperformance} shows that the Flint version of the smart contract is marginally more expensive to run in terms of gas costs. However, the Oyente and Mythril analysers find 4 potential integer overflows on lines 30 (the array can become too large), and lines 34, 43, and 48 (the integer value in the \texttt{balances} mapping can become too large). They do not find any potential integer overflows in the Flint code.

\subsection{The Solidity Attacks}

In section~\ref{solidityattacks} we described five attacks on four contracts.  In section~\ref{flintproject}, we discussed how a Flint programmer would be unlikely to create the \textsc{TheDAO}'s vulnerability. When implementing a Multi-sig wallet, Flint's initialisation can't be omitted and fallback functions cannot update state. Flint libraries are stateless but a programmer doesn't have to use them. When implementing a King of Ether contract, the Flint compiler would reject code that didn't access the result of a function. In the Proof of Weak Hands Coin contract wrap-around semantics of Solidity integers used to hold the currency, caused an overflow. Its vulnerabilities could have been prevented using Flint's {\tt{Asset}}s, in a similar way to our bank example in section~\ref{bank}. The same errors could be in a Flint contract by not using an {\tt{Asset}} for the new {\tt{coin}} and using our non-default wrap around integers  {\verb+{balanceOf[caller] = balanceOf[caller] &- amount}+}.  At this stage in Flint's development, only {\tt{Wei}} is used with {\tt{@payable}}. The plan is to extend {\tt{@payable}} to work with any currency.

As the vulnerabilities in these contracts were the starting point for the design of Flint it isn't surprising that the Flint code for them looks more robust. The Flint and Solidity code for the four (simplified) contracts is available at~\cite{flintgithubcasestudies}.

\section{Related Work}

\paragraph{\bf{Languages}}
For traditional computer architectures, languages such as Java\cite{java}, Rust\cite{rust}, and Pony\cite{pony} have been designed to prevent writing unsafe code. For instance, Java prevents direct access to memory, Rust uses \textit{ownership types} to efficiently free memory, and Pony uses \textit{reference capabilities} to prevent data races in concurrent programs. In contrast, even though the Ethereum platform requires smart contract programmers to ensure the correct behaviour of their program before deployment, Solidity lacks even a strong static type system to catch errors. Solidity provides both optional modifiers and assertions which programmers can use to decorate their contracts.

The Ethereum Foundation has created several other programming languages for writing smart contracts.
Lisp Like Language\cite{lll} (LLL)was abandoned in favour of higher level languages. Serpent\cite{serpent} is a high-level programming language with a syntax similar to Python's that was deprecated due to numerous security issues, see~\cite{zeppelinserpent}. Solidity, was an attempt to solve the issues these programming languages presented. Since Solidity, probably due to the vulnerabilities that have occurred, newer programming languages have been developed:

Vyper\cite{vyper} inspired by Python, aims at providing developers with better security and more intuitive semantics than Solidity.  Like Flint, it doesn't have contract inheritance or infinite loops. However, its dynamically type system seems inappropriate for a language for robust contract development. Like in Solidity, assertions are provided to prevent vulnerabilities. Flint, too has assertions, but we don't place as great an importance on the for vulnerability prevention as programmers can easily forget to include them. 

Like Flint with its protection blocks, the Bamboo\cite{bamboo} programming language allows reasoning about smart contracts as state machines. Developers define which functions can be called in each state, and the language provides constructs to specify changes of state explicitly. Bamboo does not present any additional features geared towards the safety of programs.

There are several large organisations developing the Linux foundation's Hyperledger Fabric\cite{fabric} framework which can run on the EVM. Contracts in this framework can be written in a variety of languages including Javascript and Go. Currently the focus seems to be more focussed on building networks and applications.

There are substantial design activities for creating new languages for Smart Contracts that do not necessarily target the Ethereum platform. These designs all have correctness as part of their design goals. Closest to Flint is Obsidian\cite{obsidian}, which includes typestate and linear types for resources. We haven't gone to full linear types for assets because we believe they would be too restrictive.  AxLang\cite{axlang} is a Scala DSL that will compile to the JVM before being targeted at the EVM.  The hope is that Scala verification tools will be useful on AxLang programs. The development of Plutus\cite{plutus} is an effort to produce an eager Haskell-like language for a range of blockchains including Ethereum. Another functional language in the design stage, Formality\cite{formality}, will target Ethereum's web assembly language eWASM\cite{ewasm}. It isn't obvious what eWASM will add to the safety of the compiler.
\paragraph{\bf{Protections}}
Our protection blocks define both typestate\cite{fickle,typestate} and caller protections. 
There have been many other techniques that protected code from being called by unwanted callers.
Solidity modifiers allow checking for \textit{any} type of assertion before the body of a function is entered. However, Solidity programmers are not required to use modifiers. As we believe protecting privileged functions in smart contracts is a basic requirement, Flint requires these to be coded before writing functions,\footnote{A programmer could decide, however, to write all functions in an \texttt{any} block with no typestate.} Our implementation is also more efficient for internal calls, as shown in section \ref{callercapseval}. Flint users can use the  \texttt{assert} function to perform other types of checks at runtime.

Ideas similar to our caller protections have arisen several times in the past. In the original definition of a capabilities by
Jack Dennis et al.\cite{dennis1966programming} they regard a capability as an unforgeable token (a number) which when possessed by a user, allows access to a resource.
In Flint, a caller protection is linked with an Ethereum address, which can be regarded as an unforgeable token. An individual has the authority to call a function if it
possesses the appropriate private key, and transferring authority is done by
simply sharing the private key. In Pony\cite{pony}, reference capabilities (similar to Flint's protections) are associated with object references and are part of the object's type. Flint's caller protections are not
encoded in the functions' type.  They are also similar to roles in role-oriented programming\cite{roles}. The underlying idea of role-oriented programming is to capture the human idea of roles and their interactions when writing code.  Operating systems may use ACLs (access control lists)\cite{access} to specify which users or system processes are granted access to which operations on objects. 
 
 \paragraph{\bf{Analysis tools}}
In addition to designing the right language there is work aimed at finding vulnerabilities, focussing on early stages in the development cycle. Oyente\cite{makingsmartcontractssmarter} and Mythril\cite{Mythril} are both dynamic analysis tools.
The Oyente symbolic execution tool analyses EVM bytecode. The authors of the tool claim they have found that 50\% of smart contracts on Ethereum's network has at least one vulnerability. Oyente finds issues such as timestamp dependencies, mishandled exceptions, and detects reentrancy calls.
Mythril uses concolic analysis to determine execution paths. Mythril tries to find a variety of vulnerabilities, such as timestamp dependencies, integer overflows, and reentrancy issues.
One of the concerns with both Oyente and Mythril is that they are not closely tied into contract development so vulnerabilities that they could catch may not be caught because the tool is not used.

The online Remix\cite{remix} IDE for Solidity has a built-in code analyser to find bug, such as reentrancy bugs and incorrect usage of low-level calls. In our testing, when running on the examples given in \ref{solidityattacks}, the Remix Analyser warned when low-level calls' return value was not checked, but was not able to find any reentrancy call issues that could be found by the dynamic analysis tools.

Converting to F*\cite{bhargavan2016formal} involves translating Solidity code to F*, decompiling EVM bytecode to F*, then checking equivalence between the two translations. Scilla\cite{sergey2018scilla} is a continuous passing style--based intermediate representation language for smart contracts that converts Scilla code to the Coq theorem prover where properties of the contract can be proven.

An SMT module has been added to the Solidity compiler\cite{smt-solidity}. Its current capabilities include catching infinite loops, arithmetic overflows, and re-entrancy, problems that by construction cannot exist in Flint code. The plans are to extend the subset of Solidity that it deals with and the problems it looks for.

\section{Conclusions and Future Work}

The goal of Flint is to make it easy to write smart contracts which are are safe by construction. Although our current compiler targets the EVM, the language design was not much influenced by  the choice of backend and could be retargeted. Flint needs to contain the constructs that programmers want in order to write easy to read smart contracts and no more. We haven't had difficulty translating existing smart contracts on the web into Flint. Omitting constructs that have been implicated many times such as sophisticated fallback functions, infinite loops, default integers with wrap-around semantics, functions that are public and mutable by default, will prevent previous errors re-appearing in new contracts. 

We have provided many features such as protection blocks, restricting access with typestate and callers, non-wraparound integers, Wei implemented as an asset, and functions private and immutable by default. Our protection blocks shift the design pattern for developing smart contracts and we believe that should be a game changer. Flint programmers will have to think about who and when each piece of code can be accessed before starting to write any individual functions. Adding who can execute code after writing it by using modifiers is likely to be far more error prone than defining the gateway code first. Vyper has omitted Solidity's new(ish) modifiers because the designers are concerned about how error prone they are.

Our language design is certainly not frozen. Wei is checked by the compiler, but we need to ensure that all assets are treated as carefully as Wei. We also would like to be able to deal with collections of assets elegantly. We also have not investigated the problems caused by aliasing. Another area of interest is interacting with non-Flint contracts as we need to be confident that bad Solidity contracts do not pollute Flint contracts. Re-entrancy can be prevented by limiting the number of times (using typestate) a foreign and a Flint contract are allowed to interact. Our performance figures are in line with those of Solidity, and this is without an optimisation pass in our compiler. We do have plans to optimise, but only after the compiler provides the full functionality.  We are currently working on a toolchain to provide the Flint programmer the toolset software engineers rely on.

Providing programmers a language with fewer ways they can make mistakes, is a first step towards ensuring  contracts will be correct. But contracts have to implement the intentions of the people who request them and programming languages do not always encode intentions. Intentions need to be captured in specifications and existing specification techniques do not capture the long-term, open-world nature of smart contracts. 

\paragraph{\bf{Acknowledgements}} We would like acknowledge Aurel Bily, Catalin Cracium, Calin Farcas, Yicheng Luo, Constantin Mueller, and Niklas Vangerow for their work on Flint, both the language and the toolchain. We would like to thank Nobuko Yoshida and Alastair Donaldson for supporting some of this work. It has been partially supported by grant EPSRC EP/K011715/1 and a grant from the Ethereum Foundation.

\bibliography{sample}
\newpage \noindent

{\bf\large{Appendices}}
\appendix
\section{The SimpleDAO}\label{simple}

\begin{lstlisting}[language=Flint, name=AppendixDAO]
// A condensed smart contract for a Decentralized Autonomous Organization (DAO)
 // to automate organizational governance and decision-making.

 // Removed features:
 // - Spliting DAO
 // - Grace/Quorum Periods
 // Moved consensus features to curator to simplify contraxt

 struct Proposal {
   var proposer: Address
   var payout: Int
   var recipient: Address
   var yea: Int = 0
   var nay: Int = 0
   var finished: Bool = false
   var success: Bool = false
   var voted: [Address: Bool] = [:]

   mutating init(proposer: Address, payout: Int, recipient: Address) {
     self.proposer = proposer
     self.payout = payout
     self.recipient = recipient
   }
 }

 contract SimpleDAO (Join, Propose, Vote) {
   var curator: Address
   visible var proposal: Int = 0
   var proposals: [Proposal] = []
   var balances: [Address: Wei] = [:]
 }

 SimpleDAO @(any) :: caller <- (any) {
   public init(curator: Address){
     self.curator = curator
     become Join
   }

   public mutating fallback() {
     fatalError()
   }

   public func tokenHolder(addr: Address) -> Bool {
     return balances[addr].getRawValue() != 0
   }

   public func getTotalStake() -> Int {
     var sum: Int = 0
     for let balance: Wei in balances {
       sum += balance.getRawValue()
     }
     return sum
   }
 }

 SimpleDAO @(Join) :: caller <- (any) {

   @payable
   public mutating func join(implicit value: inout Wei) {
     balances[caller].transfer(&value)
   }
 }

 SimpleDAO @(Join) :: (curator) {
   public mutating func joinTimeElapsed() {
     become Propose
   }
 }

 SimpleDAO @(Propose) :: caller <- (tokenHolder) {
   public mutating func newProposal(value: Int, recipient: Address) -> Int {
     // Sanity checks omitted to be concise
     let pID: Int = proposals.size + 1;
     proposals[pID] = Proposal(caller, value, recipient)
     return pID
   }

   public mutating func leave() {
     send(caller, &balances[caller])
   }
 }

 SimpleDAO @(Propose) :: (curator) {
   public mutating func beginVote(proposal: Int) {
     self.proposal = proposal
     become Vote
   }
 }

 SimpleDAO @(Vote) :: caller <- (tokenHolder) {
   public mutating func vote(approve: Bool) {
     if proposals[proposal].voted[caller] {
       fatalError()
     }

     if approve {
       proposals[proposal].yea += balances[caller].getRawValue()
     } else {
       proposals[proposal].nay += balances[caller].getRawValue()
     }

     proposals[proposal].voted[caller] = true
   }

   public mutating func executeProposal() {
     if(caller != proposals[proposal].proposer || proposals[proposal].finished) {
       fatalError()
     }

     proposals[proposal].finished = true
     // Quorum check omitted for brevity.
     if proposals[proposal].yea > proposals[proposal].nay {
        proposals[proposal].success = true
        let transfervalue: Wei = Wei(0)
        let totalstake: Int = getTotalStake()
        for let value: Wei in balances {
          let rawvalue: Int = (proposals[proposal].payout * value.getRawValue()) / totalstake
          transfervalue.transfer(&value, rawvalue)
        }
        send(proposals[proposal].recipient, &transfervalue)
     }

     become Propose
   }
 }
\end{lstlisting}
}

 \section{The Bank Contract}\label{AppendixBank}

 \begin{lstlisting}[language=Flint, name=Appendixbank]

// Contract declarations contain only their state properties.
contract Bank {
  var manager: Address
  var balances: [Address: Wei] = [:]
  var accounts: [Address] = []
  var lastIndex: Int = 0

  var totalDonations: Wei = Wei(0)

  event didCompleteTransfer {
    let from: Address
    let to: Address
    let value: Int
  }
}

// The functions in this block can be called by any user.
Bank :: account <- (any) {
  public init(manager: Address) {
    self.manager = manager
  }

  // Returns the manager's address.
  public mutating func register() {
    accounts[lastIndex] = account
    lastIndex += 1
  }

  public func getManager() -> Address {
    return manager
  }

  @payable
  public mutating func donate(implicit value: Wei) {
    // This will transfer the funds into totalDonations.
    totalDonations.transfer(&value)
  }
}

// Only the manager can call these functions.
Bank :: (manager) {

  // This function needs to be declared "mutating" as its body mutates
  // the contract's state.
  public mutating func freeDeposit(account: Address, amount: Int) {
    var w: Wei = Wei(amount)
    balances[account].transfer(&w)
  }

  public mutating func clear(account: Int) {
    balances[account] = Wei(0)
  }

  // This function is non-mutating.
  public func getDonations() -> Int {
    return totalDonations.getRawValue()
  }
}

// Any user in accounts can call these functions.
// The matching user's address is bound to the variable account.
Bank :: account <- (accounts) {
  public func getBalance() -> Int {
    return balances[account].getRawValue()
  }

  public mutating func transfer(amount: Int, destination: Address) {
    // Transfer Wei from one account to another. The balances of the
    // originator and the destination are updated atomically.
    // Crashes if balances[account] doesn't have enough Wei.
    balances[destination].transfer(&balances[account], amount)

    // Emit the Ethereum event.
    emit didCompleteTransfer(from: account, to: destination, value: amount)
  }

  @payable
  public mutating func deposit(implicit value: Wei) {
    balances[account].transfer(&value)
  }

  public mutating func withdraw(amount: Int) {
    // Transfer some Wei from balances[account] into a local variable.
    let w: Wei = Wei(&balances[account], amount)

    // Send the amount back to the Ethereum user.
    send(account, &w)
  }
}
\end{lstlisting}
 \section{The Flint Grammar}
\label{flintgrammar}

The grammar is specified in Backus-Naur form. Elements in square brackets are tokens, and elements in parentheses are optional.
\small{
\begin{verbatim}
; FLINT GRAMMAR (RFC 7405)

; TOP LEVEL
topLevelModule = 1*(topLevelDeclaration CRLF);

topLevelDeclaration = contractDeclaration
                    / contractBehaviourDeclaration
                    / structDeclaration
                    / enumDeclaration
                    / traitDeclaration;

; CONTRACTS
contractDeclaration = %s"contract" SP identifier SP [identifierGroup] SP "{" *(WSP variableDeclaration CRLF) "}";

; VARIABLES
variableDeclaration = [*(modifier SP)] WSP (%s"var" / %s"let") SP identifier typeAnnotation [WSP "=" WSP expression];

; TYPES
typeAnnotation = ":" WSP type;

type = identifier ["<" type *("," WSP type) ">"]
     / basicType
     / arrayType
     / fixedArrayType
     / dictType;

basicType = %s"Bool"
          / %s"Int"
          / %s"String"
          / %s"Address";

arrayType      = "[" type "]";
fixedArrayType = type "[" numericLiteral "]";
dictType       = "[" type ":" WSP type "]";

; ENUMS
enumDeclaration = %s"enum" SP identifier SP [typeAnnotation] SP "{" *(WSP enumCase CRLF) "}";
enumCase        = %s"case" SP identifier
                / %s"case" SP identifier WSP "=" WSP expression;

; TRAITS
traitDeclaration = %s"struct" SP %s"trait" SP identifier SP "{" *(WSP traitMember CRLF) "}"
                 / %s"contract" SP %s"trait" SP identifier SP "{" *(WSP traitMember CRLF) "}"
                 / %s"external" SP %s"trait" SP identifier SP "{" *(WSP traitMember CRLF) "}";

traitMember = functionDeclaration
            / functionSignatureDeclaration
            / initializerDeclaration
            / initializerSignatureDeclaration
            / contractBehaviourDeclaration
            / eventDeclaration;

; EVENTS
eventDeclaration = %s"event" identifer parameterList

; STRUCTS
structDeclaration = %s"struct" SP identifier [":" WSP identifierList ] SP "{" *(WSP structMember CRLF) "}";

structMember = variableDeclaration
             / functionDeclaration
             / initializerDeclaration;

; BEHAVIOUR
contractBehaviourDeclaration = identifier WSP [stateGroup] SP "::" WSP [callerBinding] callerProtectionGroup WSP "{" *(WSP contractBehaviourMember CRLF) "}";

contractBehaviourMember = functionDeclaration
                        / initializerDeclaration
                        / fallbackDeclaration
                        / initializerSignatureDeclaration
                        / functionSignatureDeclaration;

; ACCESS GROUPS
stateGroup            = "@" identifierGroup;
callerBinding         = identifier WSP "<-";
callerProtectionGroup = identifierGroup;
identifierGroup       = "(" identifierList ")";
identifierList        = identifier *("," WSP identifier)

; FUNCTIONS + INITIALIZER + FALLBACK
functionSignatureDeclaration    = functionHead SP identifier parameterList [returnType]
functionDeclaration             = functionSignatureDeclaration codeBlock;
initializerSignatureDeclaration = initializerHead parameterList
initializerDeclaration          = initializerSignatureDeclaration codeBlock;
fallbackDeclaration             = fallbackHead parameterList codeBlock;

functionHead    = [*(attribute SP)] [*(modifier SP)] %s"func";
initializerHead = [*(attribute SP)] [*(modifier SP)] %s"init";
fallbackHead    = [*(modifier SP)] %s"fallback";

attribute = "@" identifier;
modifier  = %s"public"
          / %s"mutating"
          / %s"visible";

returnType = "->" type;

parameterList = "()"
              / "(" parameter *("," parameter) ")";

parameter          = *(parameterModifiers SP) identifier typeAnnotation [WSP "=" WSP expression];
parameterModifiers = %s"inout" / %s"implicit"

; STATEMENTS
codeBlock = "{" [CRLF] *(WSP statement CRLF) WSP statement [CRLF]"}";
statement = expression
          / returnStatement
          / becomeStatement
          / emitStatement
          / forStatement
          / ifStatement;

returnStatement = %s"return" SP expression
becomeStatement = %s"become" SP expression
emitStatement   = %s"emit" SP functionCall
forStatement    = %s"for" SP variableDeclaration SP %s"in" SP expression SP codeBlock

; EXPRESSIONS
expression = identifier
           / inOutExpression
           / binaryExpression
           / functionCall
           / literal
           / arrayLiteral
           / dictionaryLiteral
           / self
           / variableDeclaration
           / bracketedExpression
           / subscriptExpression
           / rangeExpression
           / attemptExpression;

inOutExpression = "&" expression;

binaryOp = "+" / "-" / "*" / "/" / "**"
         / "&+" / "&-" / "&*"
         / "="
         / "==" / "!="
         / "+=" / "-=" / "*=" / "/="
         / "||" / "&&"
         / ">" / "<" / "<=" / ">="
         / ".";

binaryExpression = expression WSP binaryOp WSP expression;

self = %s"self"

rangeExpression = "(" expression ( "..<" / "..." ) expression ")"

bracketedExpression = "(" expression ")";

subscriptExpression = subscriptExpression "[" expression "]";
                    / identifier "[" expression "]";

attemptExpression = try expression
try = %s"try" ( "!" / "?" )

; FUNCTION CALLS
functionCall = identifier "(" [expression] *( "," WSP expression ) ")";

; CONDITIONALS
ifStatement = %s"if" SP expression SP codeBlock [elseClause];
elseClause  = %s"else" SP codeBlock;

; LITERALS
identifier = ( ALPHA / "_" ) *( ALPHA / DIGIT / "$" / "_" );
literal    = numericLiteral
           / stringLiteral
           / booleanLiteral
           / addressLiteral;

number         = 1*DIGIT;
numericLiteral = decimalLiteral;
decimalLiteral = number
               / number "." number;

addressLiteral = %s"0x" 40HEXDIG;

arrayLiteral      = "[]";
dictionaryLiteral = "[:]";

booleanLiteral = %s"true" / %s"false";
stringLiteral  = """ identifier """;
\end{verbatim}
}

\section{Assets}\label{assets}

\begin{lstlisting}[language=Flint, name=Assets]
// Any currency should implement this trait to be able to use the currency
// fully. The default implementations should be left intact, only
// `getRawValue` and `setRawValue` need to be implemented.

struct trait Asset {
  // Initialises the asset "unsafely", i.e. from `amount` given as an integer.
  init(unsafeRawValue: Int)

  // Initialises the asset by transferring `amount` from an existing asset.
  // Should check if `source` has sufficient funds, and cause a fatal error
  // if not.
  init(source: inout Self, amount: Int)

  // Initialises the asset by transferring all funds from `source`.
  // `source` should be left empty.
  init(source: inout Self)

  // Moves `amount` from `source` into `this` asset.
  mutating func transfer(source: inout Self, amount: Int) {
    if source.getRawValue() < amount {
      fatalError()
    }

    // TODO: support let _: Int = ...
    let unused1: Int = source.setRawValue(value: source.getRawValue() - amount)
    let unused2: Int = setRawValue(value: getRawValue() + amount)
  }

  mutating func transfer(source: inout Self) {
    transfer(source: &source, amount: source.getRawValue())
  }

  // Returns the funds contained in this asset, as an integer.
  mutating func setRawValue(value: Int) -> Int

  // Returns the funds contained in this asset, as an integer.
  func getRawValue() -> Int
}

struct Wei: Asset {
  var rawValue: Int = 0

  init(unsafeRawValue: Int) {
    self.rawValue = unsafeRawValue
  }

  init(source: inout Wei, amount: Int) {
    transfer(source: &source, amount: amount)
  }

  init(source: inout Wei) {
    let amount: Int = source.getRawValue()
    transfer(source: &source, amount: amount)
  }

  mutating func setRawValue(value: Int) -> Int {
    rawValue = value
    return rawValue
  }

  func getRawValue() -> Int {
    return rawValue
  }
}
\end{lstlisting}
\section{Compiler Diagnostics}
\label{diagnostics}

\noindent
{\bf{Caller Protections}}
\begin{description}
  \item Use of undeclared caller protection.\\\texttt{Caller protection 'admin' is undefined in 'Bank', or has incompatible type.} 
  \item No matching function for function call. \\\texttt{Function 'setManager' is not in scope or cannot be called using caller protection '(any)'. Note: Perhaps you meant this function, which requires caller protection '(manager)'.}
\end{description}
\noindent
{\bf{Mutation}}
\begin{description}
  \item Mutating statement in nonmutating function. \\\texttt{Use of mutating statement in a nonmutating function.}
  \item No mutating statements in mutating function (Warning).\\\texttt{Function does not have to be declared mutating: none of its statements are mutating.}
  \item Reassignment to constant. \\\texttt{Cannot reassign to value: 'manager' is a let-constant. Note: 'manager' is declared on line 18, column 12.}
\end{description}
\noindent
{\bf{Initialisation}}
\begin{description}
  \item State property is not assigned a value. \\\texttt{State property 'manager' needs to be assigned a value, as no initialiser was declared.}
	
  \item Return from initialiser without initialising all properties.\\\texttt{Return from initialiser without initialising all properties. Note: 'manager' is uninitialised.}
	
  \item Contract does not have a public initialiser.\\\texttt{Contract 'Bank' needs a public initialiser accessible using caller capability 'any'.}
    
  \item Contract has multiple public initialisers.\\\texttt{A public initialiser has already been defined. Note: A public initialiser is defined on line 5, column 6.}
    
  \item Public contract initialiser is not accessible using caller capability \texttt{any}.\\\texttt{Public contract initialiser should be callable using caller capability 'any'.}
\end{description}
\noindent
{\bf{Invalid Declarations}}
\begin{description}  
    \item Invalid redeclaration of an identifier.\\\texttt{Invalid redeclaration of 'setManager'. Note: Previous declaration on line 12, column 4.}
	
	\item Use of invalid character. The \texttt{\$} character is reserved for use in the standard library.\\\texttt{Use of invalid character '\$' in 'my\$Func'.}
	
	\item Contract Behaviour Declaration has no matching Contract Declaration.\\\texttt{Contract behaviour declaration for 'Bank' has no associated contract declaration.}
	
	\item Invalid contract behaviour declaration.\\\texttt{Contract behaviour declaration for Bank has no associated contract declaration.}
	
	\item Invalid \texttt{@payable} function.\\\texttt{receive is declared @payable but doesn't have an implicit parameter of a currency type.}
	
	\item Ambiguous @payable value parameter.\\\texttt{Ambiguous implicit payable value parameter. Only one parameter can be declared 'implicit' with a currency type.}
	
	\item Public function has a parameter of dynamic type, such as struct, array, or dictionary.\\\texttt{Function 'isSeatFree' cannot have dynamic parameters. Note: 'seat' cannot be used as a parameter.}
	
	\item Use of undeclared identifier.\\\texttt{Use of undeclared identifier 'manager'.}
	
	\item Missing return in non-void function.\\\texttt{Missing return in function expected to return 'Int'.}
	
	\item Code after return (Warning).\\\texttt{Code after return will never be executed.}
\end{description}

{\bf{Type Checking}}
\begin{description}
  \item Incompatible return type.\\\texttt{Cannot convert expression of type 'Int' to expected return type 'Address'.}
  \item Incompatible assignment.\\\texttt{Incompatible assignment between values of type Int and Wei.}
  \item Incompatible argument type.\\\texttt{Cannot convert expression of type Int to expected argument type Wei}
\end{description}

\end{document}